\documentclass[10pt, twocolumn]{extarticle}
\usepackage{graphicx}
\usepackage{hyperref}
\usepackage{amssymb}
\usepackage{amsmath}
\usepackage{amsfonts}
\usepackage{parskip}
\usepackage{titling}
\usepackage[table]{xcolor}
\usepackage[square,sort&compress,comma,numbers]{natbib}
\usepackage{multirow}
\usepackage{empheq}

\usepackage{titlesec}
\usepackage{caption}
\titleformat{\section}{\large\bf}{\thesection}{1em}{}
\titleformat{\subsection}{\bf}{\thesubsection}{1em}{}
\titleformat{\subsubsection}{\it}{\thesubsubsection}{1em}{}

\pretitle{\Large\bf\vskip 0.3em} \posttitle{\par\vskip 0.6em} \preauthor{\large} \postauthor{\par\vskip -0.0em}
\predate{\large} \postdate{\par\vskip 0.1em \rule{\linewidth}{0.5pt}} 

\newcommand{\beq}{\begin{equation}}
\newcommand{\eeq}{\end{equation}}
\newcommand{\bea}{\begin{eqnarray}}
\newcommand{\eea}{\end{eqnarray}}

\newcommand{\comment}[1]{}
\renewcommand{\d}{{\rm d}}

\begin{document}
\captionsetup{font=small}

\title{Towards the Information-Theoretic Limit of Programmable Photonics}
\author{Ryan Hamerly$^{1,2}$, Jasvith R.\ 
Basani$^3$, Alexander Sludds$^{1}$, Sri K.\ 
Vadlamani$^1$, and Dirk Englund$^1$}   
\date{\today}


\maketitle



\begin{flushleft}
\small
$^{1}$      \textit{Research Laboratory of Electronics, MIT, 50 Vassar Street, Cambridge, MA 02139, USA} \\
$^{2}$      \textit{NTT Research Inc., Physics and Informatics Laboratories, 940 Stewart Drive, Sunnyvale, CA 94085, USA} \\
$^{3}$      \textit{Department of Electrical and Computer Engineering, Institute for Research in Electronics and Applied Physics, and Joint Quantum Institute, University of Maryland, College Park, MD 20742, USA}
\end{flushleft}

{\bf\small The scalability of many programmable photonic circuits is limited by the $2\pi$ tuning range needed for the constituent phase shifters.  To address this problem, we introduce the concept of a phase-efficient circuit architecture, where the average phase shift is $\ll 2\pi$.  We derive a universal information-theoretic limit to the phase-shift efficiency of universal multiport interferometers, and propose a ``3-MZI'' architecture that approaches this limit to within a factor of $2\times$, approximately a $10\times$ reduction in average phase shift over the prior art, where the average phase shift scales inversely with system size as $O(1/\sqrt{N})$.  For non-unitary circuits, we show that the 3-MZI saturates the theoretical bound for Gaussian-distributed target matrices.  Using this architecture, we show optical neural network training with all phase shifters constrained to $\lesssim 0.2$~radians without loss of accuracy.  
}

\rule{\linewidth}{0.5pt}

Large-scale programmable photonic circuits are a key enabler to many emerging applications of optics, including quantum information processing \cite{bartolucci2023fusion, madsen2022quantum}, machine learning acceleration \cite{shen2017deep, demirkiran2023electro, xiao2021large, basani2024all}, signal processing \cite{huang2021silicon, Perez2017}, and switching \cite{seok2019wafer, harris2022passage, urata2022mission}.  The advent of reliable photonic integration, enabled by recent advances in foundry-scale silicon photonics processes \cite{shekhar2023silicon}, in principle opens the door to scaling circuits to ever-larger sizes, following an exponential growth trend analogous to Moore's Law.  However, the {\it programmablility} of this photonic circuit imposes unique physical limits to scaling that arise from the circuit's key programmable photonic element: the phase shifter.  Unlike transistors, phase shifters cannot be scaled down arbitrarily \cite{dennard1974design}, as their design is subject to tradeoffs between loss, footprint, power, speed, voltage, crosstalk, yield, and optical bandwidth
.  For example, typical thermo-optic shifters draw a static power of $P_\pi \approx 20~\text{mW}$ \cite{Harris2014, Watts2013}; reducing this $P_\pi$ comes the cost of decreased speed \cite{murray2015dense} or increased size and insertion loss \cite{Qiu2020}.  Alternative technologies such as phase-change material (PCM) and metal-oxide-semiconductor capacitor (MOSCAP) shifters draw negligible power, but incur a phase-dependent optical loss $\alpha_\pi \sim 1~\text{dB}$ \cite{takenaka2019iii, cheung2022ultra, liang2022energy}.  In either case, the high insertion loss and/or power consumption of circuits with many phase shifters poses a practical limit to scaling up programmable photonics, and leads to realistic circuits being 10--100$\times$ larger than the theoretical $O(\lambda/n)$ limit set by the wavelength of light.

The tradeoff above arises from the need for full $[0, 2\pi)$ 
programmability of every phase shifter; by contrast, implementing small phase shifts is fairly easy.  This raises the question of whether we can circumvent the tradeoff by doing ``more with less'': creating architectures that minimize the {\it average} phase shift per element in a large circuit.  In many specific cases, this is possible; for example, crossbar-based circuit switches can function with only a fraction $O(1/N)$ of phase shifters active at a given time \cite{Suzuki2018}.  Likewise, sparsity and pruning \cite{basani2023self, yu2023heavy} in interferometer meshes is possible (in specific cases) without a degradation in fidelity.  However, these examples are exceptions that prove the norm: for the more general class of {\it universal} multiport interferometers, i.e.\ circuits that implement a fully programmable unitary map between $N$ input- and output-modes $\vec{x} \rightarrow U \vec{x}$ (Fig.~\ref{fig:f1}(a)), existing architectures require that all phase shifters be active, with an average phase shift of $O(1)$ per component.

In this paper, we show that universal circuits can also economize phase shift, and provide a prescription to reduce the average phase shift per component by a factor of $\sqrt{N}$.  First, we use information theory to derive a fundamental lower bound to the average phase shift required for universality.  This bound, which scales as $O(1/\sqrt{N})$, applies to any unitary photonic architecture.  Second, we show that although the standard construction of a Mach-Zehnder interferometer (MZI) mesh \cite{reck1994experimental, clements2016optimal} leads to an average phase shift of $O(1)$, one can recover the $O(1/\sqrt{N})$ scaling with a 3-splitter design (3-MZI) \cite{hamerly2022asymptotically}, which yields a $10\times$ reduction in average phase shift for realistic circuit sizes, and approaches the information-theoretic bound to within a factor of $\approx 2\times$.  Moreover, we show that the situation is even better for non-unitary meshes: for a target distribution of Gaussian matrices, combining the diamond mesh \cite{Blass1960, Mosca2002, Taballione2019} with the 3-MZI saturates the bound.  Finally, as a practical example, we perform phase-constrained training on an optical neural network, showing minimal loss in accuracy even when bounding all phase shifts to $\lesssim 0.2$ radians.  
By treating phase shift as a resource to be economized, this work establishes a new paradigm in the design and analysis of photonic circuits, leading to new architectures that are provably optimal, and provide new headroom for increasing the circuit sizes in programmable photonics to the very-large-scale domain required for emerging applications.

\section{Information-Theoretic Phase Bound}
\label{sec:bound}

This section proves an information-theoretic lower bound to the average phase shift $\psi$ in multiport interferometers.  To do so, we begin in Sec.~\ref{sec:infdef} by defining the system mathematically, representing any lumped-element linear photonic circuit as a product of well-defined matrices, and clarifying the meaning of ``average phase shift'' in terms of moments.  With these definitions in place, Sec.~\ref{sec:derbnd} derives a lower bound on the entropy of the distribution $P(\psi)$, which can be translated into bounds on the moments through Lagrange optimization.

\subsection{Definitions}
\label{sec:infdef}

First, we formalize the problem.  To do so, we must (1) develop a general model for multiport interferometers with discrete phase shifters $\psi$, (2) define the average phase shift in terms of of the probability distribution $P(\psi)$, and (3) set the distribution of target matrices $U$.

\begin{figure}[t!]
\begin{center}
\includegraphics[width=1.00\columnwidth]{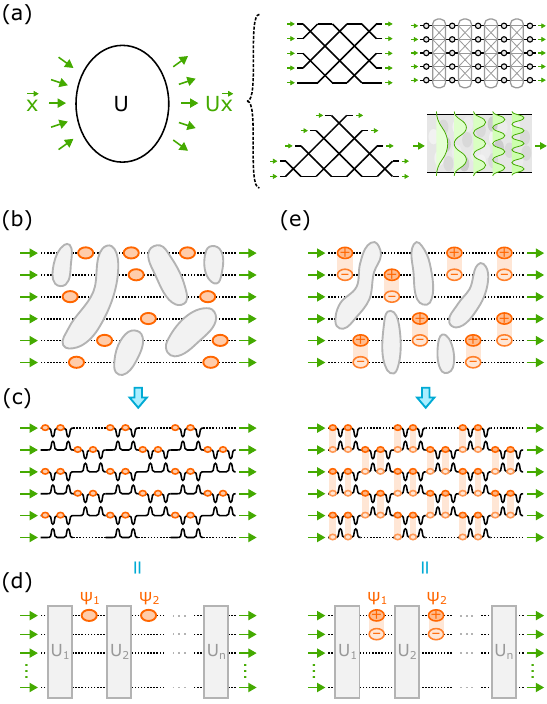}
\caption{(a) Schematic of a generic multiport interferometer, with four concrete implementations: Reck triangle, Clements rectangle, MPLC, and programmable MMI (clockwise from bottom left).  (b) Representation of a programmable circuit consisting of discrete phase shifters (colored) and fixed coupler unitaries (gray), with (c) the corresponding design of a Clements mesh, and (d) the mathematical description of the unitary as a product of phase-shift $D(\psi_i)$ and coupler $U_i$ matrices.  (e) Related push-pull construction for mesh-like multiport interferometers.}
\label{fig:f1}
\end{center}
\end{figure}

{\it 1.\ Interferometer Model---}A multiport interferometer is defined as a linear photonic circuit with $N$ input- and output-modes (Fig.~\ref{fig:f1}(a)).  When using lumped phase elements, any such circuit can be represented as a directed graph (Fig.~\ref{fig:f1}(b)) containing programmable nodes (phase shifters, colored) and fixed unitary elements (gray).  For example, a Clements mesh \cite{clements2016optimal} consists of a regular 2D array of phase shifters (Fig.~\ref{fig:f1}(c)), interleaved with $2\times 2$ splitters, but such regularity is not necessary for our proof.  The coupling matrix for any $N\times N$ circuit with $n$ phase shifters can be written as follows
\beq
	U = U_n \ldots U_2 \underbrace{\begin{bmatrix} e^{i\psi_2} & \\ & I_{N-1}\end{bmatrix}}_{D(\psi_2)} U_1 \underbrace{\begin{bmatrix} e^{i\psi_1} & \\ & I_{N-1}\end{bmatrix}}_{D(\psi_1)} U_0, \label{eq:u1}
\eeq
i.e.\ as a product of fixed unitaries $U_i$ interleaved with tunable diagonal matrices $D(\psi_i)$ representing the phase shifters (Fig.~\ref{fig:f1}(d)).

\begin{table*}[t]
\begin{center}
\begin{tabular}{c|rlrl|c}
\hline\hline
{\bf Norm} & \multicolumn{2}{c}{\bf Definition} & \multicolumn{2}{c|}{\bf System FoM} & {\bf Lower Bound} \\ \hline
$L_1$ & $\lVert\psi\rVert_1$\!\!\!\!\! & $ = \langle \sum_i |\psi|_i \rangle$ & 
      Power: \!\!\!\!\! & $P_{\rm tot} = P_\pi \langle \psi\rangle_1 /\pi$ & $0.98/\sqrt{N}$ \\
& & & Loss: \!\!\!\!\! & $\alpha_{\rm tot} = \alpha_\pi \lVert \psi \rVert_1 / \pi N$ & \\ \hline
$L_2$ & $\lVert\psi\rVert_2$\!\!\!\!\! & $ = \langle \sum_i (\psi_i)^2 \rangle^{1/2}$ &  
      Error: \!\!\!\!\! & $\mathcal{E} = \alpha_\pi \lVert\psi\rVert_2/2\pi \sqrt{N}$ & $1.28/\sqrt{N}$ \\ \hline
$L_\infty$ & $\lVert\psi\rVert_\infty$\!\!\!\!\! & $ = \text{max}_{i,U}|\psi_i(U)|$ & 
      Voltage: \!\!\!\!\! & $V_{pp} = (2/\pi) \lVert\psi\rVert_\infty V_\pi L / L$ & $2.65/\sqrt{N}$ \\ 
& & & Length: \!\!\!\!\! & $L = (2/\pi) \lVert\psi\rVert_\infty V_\pi L / V_{pp}$ & \\ \hline
IQR & $\Delta\psi_{\rm IQR}$:\!\!\!\!\! & IQR of $P(\psi)$ & \multicolumn{2}{c|}{--} & --\\
\hline\hline
\end{tabular}
\caption{Comparison between the $L_1$, $L_2$, and $L_{\rm infty}$ norms and IQR, showing the related system figures of merit governed by each norm and lower bounds derived in Sec.~\ref{sec:derbnd}}
\label{tab:t1}
\end{center}
\end{table*}

As we will see later, it is also useful to define a push-pull construction for mesh-like multiport interferometers consisting of $2\times 2$ blocks (Fig.~\ref{fig:f1}(e)), where the phase-shifter pairs are differentially driven.  Push-pull operation in an MZI can achieve the same relative phase shift with a $2\times$ reduction in footprint
and it can be shown that 
a mesh based on push-pull phase-shifter pairs is also universal and has the same phase-shift statistics.  Here, the coupling matrix is given by:
\begin{align}
	U & = U_n D_p(\psi_n) \ldots U_2 D_p(\psi_2) U_1 D_p(\psi_1) U_0, \nonumber \\
	& D_p(\psi) \equiv \begin{bmatrix} e^{i\psi/2} &  & \\ & e^{-i\psi/2} & \\ & & I_{N-2}\end{bmatrix} \label{eq:pp}
\end{align}%
{\it 2.\ Average Phase Shift---}There are multiple ways to define the average phase shift, which affect the practical performance of a circuit in different ways.  The most important are the $L_1$, $L_2$, and $L_\infty$ norms, moments of the probability distribution $P(\psi)$ for all phase shifters $\psi_i$ in the circuit, under a given distribution of target matrices $U$:
\begin{enumerate}
	\item The $L_1$ norm $\lVert\psi\rVert_1 = \langle\sum_i |\psi_i|\rangle$ is the sum of all phase shifts $|\psi_i|$, with $\langle \ldots \rangle$ denoting an average over the distribution of target unitaries.  This is closely related to the $L_1$-average phase shift $\langle \psi \rangle_1 = \lVert\psi\rVert_1/N_{\rm ph}$, which is just the $L_1$ norm divided by the total number of phase shifters.  When using thermo-optic phase shifters, the average power dissipation is given by $P_{\rm tot} = P_\pi \lVert\psi\rVert_1 /\pi$ ($P_\pi \langle \psi\rangle_1 /\pi$ per phase shifter).  The $L_1$ metric is also relevant for lossy phase-shifter platforms where $\alpha \propto \psi$, as it sets the average loss per phase shifter $\langle\alpha\rangle = \alpha_\pi \langle \psi\rangle_1 / \pi$, and in unitary mesh architectures, the total loss is given by $\alpha_{\rm tot} = \alpha_\pi \lVert \psi \rVert_1 / \pi N$, where $N$ is the number of input- and output-ports.
	\item The $L_2$ norm $\lVert\psi\rVert_2 = \langle\sum_i |\psi_i|^2\rangle^{1/2}$ shows up in unitary meshes with lossy phase shifters, as it limits a mesh's fidelity by introducing uncorrectable matrix errors due to phase-shifter loss \cite{hamerly2022accurate}.  Each lossy phase shifter leads to a path-dependent loss and a matrix error with magnitude $\lVert \Delta U \rVert = \alpha_\pi |\psi|/2\pi$, which is uncorrectable because it deviates from the manifold of unitary matrices.  These errors add in quadrature, leading to an overall matrix error of $\mathcal{E} \equiv \langle \lVert \Delta U \rVert\rangle/\sqrt{N} = \alpha_\pi \lVert\psi\rVert_2/2\pi \sqrt{N}$.
	\item The $L_\infty$ norm $\lVert\psi\rVert_\infty = \text{max}_{i,U}|\psi_i(U)|$ is the maximum phase shift for all phase shifters $\psi_i$, over all unitaries $U$ in the target distribution.  This quantifies the total phase-shift range $\Delta\psi = 2\lVert\psi\rVert_\infty$ needed to realize {\it all} target matrices, as $\psi_i \in [-\lVert\psi\rVert_\infty, +\lVert\psi\rVert_\infty]$ for all phase shifters.  This metric is especially relevant for Pockels-based phase shifters, where the voltage swing $V_{pp} = \Delta\psi\, V_\pi/\pi$ is limited by driver electronics; the $L_\infty$ norm sets a lower bound for the length of a Pockels phase shifter given this constraint: $L = (2/\pi) \lVert\psi\rVert_\infty V_\pi L / V_{pp}$.
	
	We will find that for most architectures, $\lVert\psi\rVert_{\infty} = \pi$, i.e.\ there will always be some target matrix that requires the full range $[-\pi, +\pi]$ of phase shift.  In light of this, we introduce a related figure of merit, the interquartile range $\Delta\psi_{\rm IQR}$.
\end{enumerate}
These definitions are summarized and contrasted in Table~\ref{tab:t1}.  

{\it 3.\ Target Matrix Distribution---}All of these metrics depend on the distribution $P(\psi)$, which in turn depends on the distribution of target matrices.  The target matrix distribution is essential to any discussion of ``phase-optimal'' mesh architectures, and is very problem-dependent.  Here, because our focus is on {\it universal} circuits, we consider $U$ sampled uniformly over all matrices with respect to the Haar measure, which is an invariant measure for all matrices in $U(N)$ \cite{Haar1933, Tung1985}; this distribution has been widely used in previous works to study the statistical properties of universal multiport interferometers \cite{Russell2017, burgwal2017using, Pai2019, fldzhyan2020optimal, hamerly2022accurate, hamerly2022stability, hamerly2022asymptotically}.

\subsection{Deriving the Bound}
\label{sec:derbnd}

\begin{figure}[b!]
\begin{center}
\includegraphics[width=1.00\columnwidth]{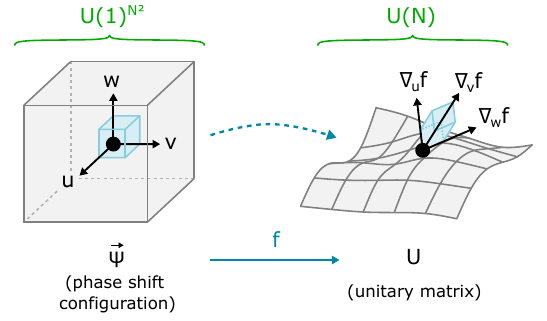}
\caption{Visualization of the map $f: \vec{\psi} \rightarrow U$ implemented by a multiport interferometer, where the unit volume $\d V_\psi$ is mapped to a parallelepiped spanned by the gradient vectors $\nabla_{\psi_m} f$ and has a volume $\d V_U = |\partial(U)/\partial(\psi)| \d V_\psi$.
}
\label{fig:f2}
\end{center}
\end{figure}

With these definitions in place, deriving the information-theoretic bound is a two-step process: (1) we show that the map between phase shifts and matrices is contractive, and use this property to lower-bound the entropy of the distribution over $\psi$, and (2) given this entropy bound, we use Lagrange multipliers to find an upper bound for the moments of $P(\psi)$.

Start by considering the function $f: U(1)^{N^2} \rightarrow U(N)$ that maps a vector of phase shifts $\vec{\psi}$ to a unitary matrix $U$ (Fig.~\ref{fig:f2}).  We can show that $f$ is contractive by recalling (Eq.~(\ref{eq:u1})) that the dependence of $U$ on any variable $\psi_m$ takes the form $U = U_{\rm post} D(\psi_m) U_{\rm pre}$.  Since $U_{\rm pre}$ and $U_{\rm post}$ are unitary, any small change to $\psi_m$ perturbs the matrix by $\lVert \Delta U\rVert = \lVert D'(\psi_m) \rVert \Delta\psi_m = \Delta\psi_m$.  This implies that the gradient vectors are all normalized, i.e.\ $|\nabla_{\psi_m} f| = 1$ along all directions $\psi_m$, so $f$ is contractive, with determinant of the Jacobian bounded by:
\beq
	\Bigl| \frac{\partial(U)}{\partial(\psi)}\Bigr|  \leq 1
\eeq
with equality holding iff the gradients $|\nabla_{\psi_m} f| = 1$ are orthogonal.  Now consider the probability distribution $P(U)$ of unitaries and its pullback (via $f$) to the space of phase shifts $P(\psi)$: the information entropies of these two distributions are related by the log-average contraction factor $H_U = H_\psi + \langle \log|\partial(U)/\partial(\psi)| \rangle$.  A contractive map reduces the information entropy, $H_U \leq H_\psi$, since $\log|\partial(U)/\partial(\psi)| \leq 1$.  Given that we have chosen to sample target matrices uniformly over the Haar measure, the entropy $H_U$ is directly related to the volume of $U(N)$ \cite[p.~194]{lando2004graphs}, giving us a lower bound for $H_\psi$:
\beq
	H_\psi \geq H_U = \log[\text{vol}(U(N))] \approx \frac{N^2}{2} \log\Bigl(\frac{2\pi e^{3/2}}{N}\Bigr) \label{eq:hu}
\eeq
This entropy bound can be translated into a bound on the {\it moments} of phase shifts $\psi$ using the method of Lagrange multipliers.  For simplicity, consider first the case of a single-variable distribution $p(x)$ with entropy $H/N^2$ (for the multivariate case $P(\vec{\psi})$, the moments are minimized when entropy is equipartitioned among all $N^2$ modes).  For any positive weighting function $w(x)$, we find the bound on the weighted average $\int{p(x)w(x)\d x}$ by solving the constrained optimization problem
\beq
	\min_p  \int{p(x) w(x) \d x}\ \ \ \text{s.t.}\ \begin{cases}
		\int{p(x)\d x} = 1 \\
		\int{p(x)\log p(x)\d x} \leq -H/N^2 \end{cases}
\eeq 
where the constraints impose normalization and the entropy bound, and the form of $w(x)$ encodes the moment we wish to minimize, as discussed below.  This problem has the Lagrange dual
\beq
	\nabla_{\lambda,\mu} \Bigl(\min_p \int{p(x)\bigl[w(x) + \lambda + \mu \log p(x)\bigr] \d x}\Bigr) = 0 \label{eq:lv}
\eeq
The minimization in Eq.~(\ref{eq:lv}) can be obtained analytically: it gives the solution $p(x) = \exp\bigl(-(w(x)+\lambda+\mu)/\mu\bigr)$ where $(\lambda, \mu)$ are set to satisfy the constraints.  To find the bound on the $L_p$ norm $\langle x \rangle_p = (\int{x^p p(x)\d x})^{1/p}$, we set $w(x) = x^p$; the minimization yields a distribution of the form
\beq
	p(x) \propto e^{-|x|^{p}/\mu} \rightarrow 
	\begin{cases}
		e^{-|x|/a} & p = 1 \\
		e^{-x^2/2\sigma^2} & p = 2 \\
		H(x_{\rm max}-|x|) & p = \infty
	\end{cases}
\eeq
i.e.\ an exponential, Gaussian, and uniform distribution for $p = 1, 2$, and $\infty$, respectively, with moments
\beq
	\langle x \rangle_p = \frac{e^{H/N^2-1/p}}{2p^{1/p} \Gamma(1+1/p)} \rightarrow
	\begin{cases}
		e^{H/N^2}/2e & p = 1 \\
		e^{H/N^2}/\sqrt{2\pi e} & p = 2 \\
		e^{H/N^2}/2 & p = \infty 
	\end{cases} \label{eq:psibnd0}
\eeq
With the minimum entropy per mode given by $H_U/N^2 = \tfrac12 \log(2\pi e^{3/2}/N)$ under Eq.~(\ref{eq:hu}), we find the following bounds for the moments:
\begin{align}
	\langle \psi \rangle_1 & \geq \underbrace{\sqrt{\frac{\pi}{2e^{1/2} N}}}_{\approx\, 0.98/\sqrt{N}}, 
	& \langle \psi \rangle_2 & \geq \underbrace{\sqrt{\frac{e^{1/2}}{N}}}_{\!\!\!\approx\, 1.28/\sqrt{N}\!\!\!},
	& \langle \psi \rangle_\infty & \geq \underbrace{\sqrt{\frac{\pi e^{3/2}}{2N}}}_{\!\!\approx\, 2.65/\sqrt{N}\!\!} \label{eq:psibnd}
\end{align}
For MZI-based circuits, the bound is stricter, since any MZI-based circuit can be represented using the push-pull form in Eq.~(\ref{eq:pp}), plus a diagonal matrix for the phase screen $\text{diag}(\phi_1, \ldots, \phi_N)$.  With respect to every MZI phase shift $\psi_m$, the unitary $U$ has the form $U = U_{\rm post} D_p(\psi_m) U_{\rm pre}$, and perturbations lead to a matrix change $\lVert \Delta U \rVert = \Delta\psi_m/\sqrt{2}$ that is a factor of $\sqrt{2}$ smaller.  Since there are $N(N-1)$ phase-shifter pairs in the mesh, this leads to stricter bound on the Jacobian $|\partial(U)/\partial(\psi)| \leq 2^{-N(N-1)/2}$, and a correspondingly larger entropy bound $H_\psi \geq H_U + \tfrac12 N(N-1)\log(2) \approx (N^2/2)\log(4\pi e^{3/2}/N)$, i.e.\ $H/N^2$ increases by $\tfrac12\log(2)$.  Given the $e^{H/N^2}$ dependence of the bounds in Eq.~(\ref{eq:psibnd0}), this in turn increases the bounds for all moments by a factor of $\sqrt{2}$.

\section{Reaching the Bound with MZI Meshes}

The upshot of the previous section is that for any $N\times N$ unitary circuit, universality requires a minimum average phase shift that scales as $O(1/\sqrt{N})$.  How well can we reach this limit in actual devices?  We will show in this section that the standard architectures---triangular \cite{reck1994experimental} or rectangular \cite{clements2016optimal} meshes of MZIs---are quite far from the bound, not even enjoying the $O(1/\sqrt{N})$ scaling.  However, by replacing the MZI with a 3-MZI as proposed in Ref.~\cite{hamerly2022asymptotically}, we can restore this scaling and come within a factor of $2\times$ of the bound.  For realistic mesh sizes, this corresponds to a $\approx 10\times$ reduction in the average phase shift compared to the standard MZI mesh.

We start by reviewing the theory of programmable meshes (Sec.~\ref{sec:genmzi}), which allows us to calculate the phase-shift distribution, assuming a Haar-uniform distribution of target matrices \cite{Russell2017}; given this distribution, we obtain general formulas for the phase-shift moments.  Next, we apply this theory to the MZI (Sec.~\ref{sec:mzi}) and 3-MZI (Sec.~\ref{sec:3mzi}) meshes and obtain analytic expressions for the moments, which scale as $O(1)$ and $O(1/\sqrt{N})$, respectively.  Finally, we validate these results with a comparison to numerical simulations (Sec.~\ref{sec:num}).

\subsection{General Theory}
\label{sec:genmzi}

\begin{figure}[tbp]
\begin{center}
\includegraphics[width=1.00\columnwidth]{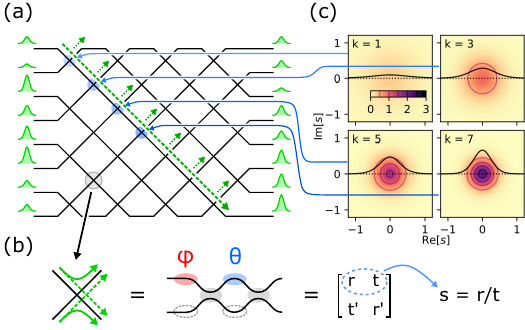}
\caption{(a) $8\times 8$ Clements mesh, which consists of a rectangular array of tunable crossings (the phase screen is omitted for clarity).  (b) MZI crossing and its representation by a $2\times 2$ matrix $T$ (Eq.~(\ref{eq:tmat})) with splitting ratio $s$.  (c) $P(s)$ for MZIs of rank 1, 3, 5, and 7 for Haar-uniform unitaries, showing how the distribution becomes concentrated as one moves to the center of the mesh.}
\label{fig:f3}
\end{center}
\end{figure}

Recall that a beamsplitter mesh \cite{reck1994experimental, clements2016optimal} is constructed from $2\times 2$ crossings (Fig.~\ref{fig:f3}(a)), each represented by a $2\times 2$ block matrix
\beq
	T \equiv \begin{bmatrix} r & t \\ t' & r' \end{bmatrix}, \label{eq:tmat}
\eeq
where $r, t, r', t'$ depend on the crossing's geometry and phase shifter settings $(\theta, \phi)$.  Every crossing has a complex-valued {\it splitting ratio} $s = r/t \in \mathbb{C}$, and to program the mesh to realize matrix $U$, we set the splitting ratios to target values given by the Givens rotations that diagonalize $U$ \cite{hamerly2022accurate}.  The distribution of these target $s$, calculated in Ref.~\cite{Russell2017} for Haar-random unitaries
\beq
	P_k(s) = \frac{k}{\pi(1+|s|^2)^{k+1}} \stackrel{k \gg 1}{\longrightarrow} (k/\pi) e^{-(k+1)|s|^2} \label{eq:ps}
\eeq
depends on the location of the crossing through the rank index $k$ (four examples are shown in Fig.~\ref{fig:f3}(b); $k$ increases as one moves to the mesh interior).  As meshes become large, a majority of the crossings have $k \gg 1$ and the $P_k(s)$ clusters near the cross state $s = 0$, corresponding to weak scatterers $|s| \approx |r| \ll 1$.  This makes sense, as weak scattering is necessary to enable ``ballistic'' transport of photons down diagonals (as opposed to diffusive transport \cite{Pai2019}), allowing all-to-all connectivity in the realized unitary.

A universal $N\times N$ unitary mesh has $N^2$ phase shifters: $\tfrac12N(N-1)$ $\theta$ and $\phi$ phase-shifters in the tunable crossings (whose values are determined from the splitting ratios distributed by Eq.~(\ref{eq:ps})), and $N$ phase-shifters in the external phase screen (whose phase shifts are uniformly distributed over $[-\pi, +\pi]$ for Haar-random unitaries).  To get the moments, we must evaluate quantities of the form $\langle f(\psi) \rangle$, which are averages over all phase shifters in the mesh.  These can be expressed in terms of a weighted average of the $\theta$, $\phi$, and phase-screen contributions:
\beq
	\langle f(\psi) \rangle = \frac{1}{N^2} \Bigl( \frac{N(N-1)}{2} \bigl(\langle f(\theta) \rangle + \langle f(\phi) \rangle \bigr) + N \langle f(\psi) \rangle_{\rm screen} \Bigr) \label{eq:fpsi1}
\eeq
where the $\theta, \phi$ averages are the sum over all crossings, indexed by rank $k$, recalling that there are $N-k$ crossings of rank $k$ in a size-$N$ mesh \cite{Russell2017}.  For example, for $\langle f(\theta) \rangle$ we would have:
\beq
	\langle f(\theta) \rangle = \frac{2}{N(N-1)} \sum_k (N-k) \langle f(\theta)\rangle_k \label{eq:fsumk}
\eeq
Following Eq.~(\ref{eq:fpsi1}), the $L_1$ moment $\langle \psi\rangle_1 = \langle |\psi|\rangle$ is the average of the $\theta$ and $\phi$ moments, plus an $O(1/N)$ correction for the phase screen:
\beq
	\langle \psi \rangle_1 \rightarrow \frac{\langle \theta \rangle_1 + \langle \phi \rangle_1}{2} + O(1/N) \label{eq:psi1}
\eeq
Since this moment will always be $\gtrsim 1/\sqrt{N}$ under the information entropy arguments of Sec.~\ref{sec:bound}, we can ignore the $O(1/N)$ term in the large-$N$ limit.

For the $L_2$ moment $\langle \psi\rangle_2 = \langle \psi^2 \rangle^{1/2}$, the phase screen in Eq.~(\ref{eq:fpsi1}) cannot be ignored.  Due to the uniform distribution, $\langle \psi^2 \rangle_{\rm screen} = \pi/3$, which gets combined with the $\theta$ and $\phi$ terms, when they are added in quadrature.  In the large-$N$ limit, we have:
\beq
	\langle \psi \rangle_2 \rightarrow \sqrt{\frac{(\langle \theta \rangle_2)^2 + (\langle \phi \rangle_2)^2}{2} + \frac{\pi}{3N}} \label{eq:psi2}
\eeq
As before, the phase screen manifests as an additional $O(1/N)$ term.  In this case, however, the term appears in quadrature summation, so it can still influence the final result if the other averages are $\langle \theta \rangle_2 \sim \langle \phi \rangle_2 \sim O(1/\sqrt{N})$, as we expect from the Sec.~\ref{sec:bound} bounds.  Therefore, we must include this term in the $L_2$ average.

The $L_\infty$ moment is always $\langle \psi \rangle_\infty = \pi$ due to the phase screen, making it a poor figure of merit to compare mesh types.  The related IQR figure is found by obtaining the total probability density in an analogous manner
\begin{align}
	P(\psi) & = \frac{1}{N^2} \Bigl( \frac{N(N-1)}{2} \bigl(P(\theta) + P(\phi) \bigr) + N P_{\rm uniform}(\psi)\Bigr) \nonumber \\
	& \begin{cases}
		P(\theta) = \frac{2}{N(N-1)} \sum_k (N-k) P_k(\theta) \\
		P(\phi) = \frac{2}{N(N-1)} \sum_k (N-k) P_k(\phi)
	\end{cases}
\end{align}
and computing the IQR of $P(\psi)$.

\subsection{MZI Mesh}
\label{sec:mzi}

\begin{figure}[t!]
\begin{center}
\includegraphics[width=1.00\columnwidth]{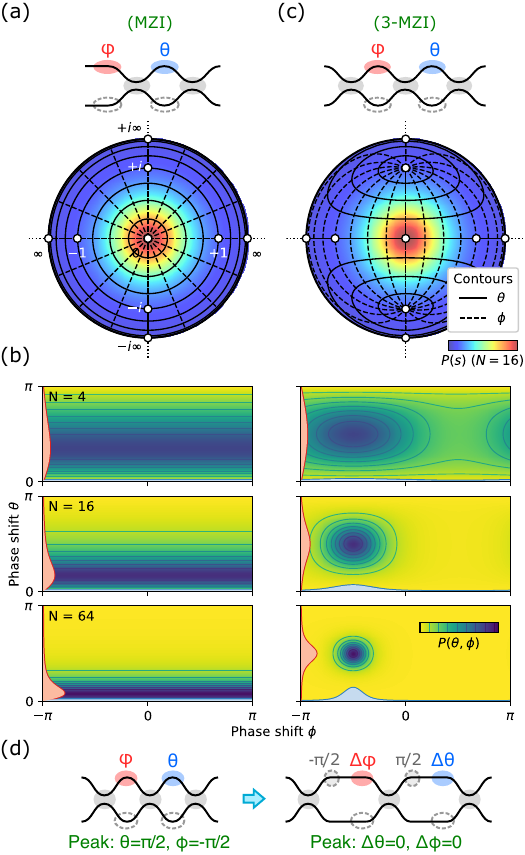}
\caption{(a) Standard MZI crossing, which maps $(\theta, \phi) \rightarrow s$ via polar coordinates.  The probability $P(s)$ for an $N = 16$ mesh is plotted to show the concentration near $s = 0$.  (b) Probability $P(\theta, \phi)$ as a function of mesh size, showing confinement in $\theta$ but not $\phi$.  (c) 3-MZI crossing, whose mapping is locally Cartesian near $s = 0$ and which confines both $\theta$ and $\phi$.  (d) Use of fabrication-induced phase offsets to shift the distribution $P(\theta, \phi)$ to center around zero.}
\label{fig:f4}
\end{center}
\end{figure}

Applying the above results, we now derive the average phase shift for the conventional MZI mesh.  The conventional mesh implements the tunable coupler with an MZI, with $\theta$ placed inside the interferometer, while $\phi$ adjusts the phase of the top input (Fig.~\ref{fig:f4}(a)).  This coupler realizes the following matrix
\begin{align}
	T_{\text{MZI}} & = \frac{1}{2} 
		\begin{bmatrix} 1 & i \\ i & 1 \end{bmatrix}
		\begin{bmatrix} e^{i\theta} & 0 \\ 0 & 1 \end{bmatrix} 
		\begin{bmatrix} 1 & i \\ i & 1 \end{bmatrix}
		\begin{bmatrix} e^{i\phi} & 0 \\ 0 & 1 \end{bmatrix} \nonumber \\
	  & = i e^{i\theta/2} \begin{bmatrix} e^{i\phi} \sin(\theta/2) & \cos(\theta/2) \\
	  		e^{i\phi} \cos(\theta/2) & -\sin(\theta/2) \end{bmatrix},
\end{align}
with a splitting ratio $s \equiv r/t = e^{i\phi} \tan(\theta/2)$.  The mapping $(\theta, \phi) \rightarrow s$ is a polar one, with the internal phase shifter setting the amplitude $|s| = \tan(\theta/2)$, while the external one sets the phase $\text{arg}(s) = \phi$.  Given the distribution Eq.~(\ref{eq:ps}), we can see that $\phi \in [-\pi, +\pi]$ is uniform, while $\theta$ clusters around the origin following \cite{Russell2017}:
\beq
	P_k(\theta) = k\sin(\theta/2)\cos(\theta/2)^{2k-1} \stackrel{k\gg1}{\longrightarrow} \frac{k\theta}{2} e^{-k\theta^2/4}
\eeq
This distribution is reflected in the probability density plotted in Fig.~\ref{fig:f4}(c), which is tightly confined in $\theta$, but not in $\phi$.

As a result, for large meshes $N \gg 1$, the average phase shift is entirely determined by the $\phi$ phase shifters (see Appendix \ref{sec:mzith} for a calculation of the $\theta$ moments, all of which scale subdominantly as $O(1/\sqrt{N})$).  The moments of the uniform distribution give us $\langle \phi \rangle_1 = \pi/2$, $\langle \phi \rangle_2 = \pi/\sqrt{3}$, and $\Delta\phi_{\rm IQR} = \pi$.  The average phase shift includes both internal and external phase shifters but since the former are tightly clustered around $\theta = 0$, in the large-$N$ limit the average is approximately:
\begin{align}
	\langle\psi\rangle_1 & = \frac{\langle\theta\rangle_1 + \langle\phi\rangle_1}{2} = \pi/4 \label{eq:psi1mzi} \\
	\langle\psi\rangle_2 & = \sqrt{\frac{(\langle\theta\rangle_2)^2 + (\langle\phi\rangle_2)^2}{2}} = \pi/\sqrt{6} \label{eq:psi2mzi} \\
	\Delta\psi_{\rm IQR} & = \pi/2 \label{eq:iqrmzi}
\end{align}
As expected, under every metric, the average phase shift for the MZI mesh is $O(1)$, which differs from the information-theoretic bound Eq.~(\ref{eq:psibnd}) by a scaling factor that increases with the size of the circuit.

\subsection{3-MZI Mesh}
\label{sec:3mzi}

\begin{table*}[t!]
\begin{center}
\begin{tabular}{rl|cc|cc|c}
\hline\hline
\multicolumn{2}{c|}{\bf Norm} & \multicolumn{2}{c|}{\bf Mesh} & \multicolumn{2}{c|}{\bf Bound} & {\bf 3-MZI/Bound} \\ 
& & MZI & 3-MZI & General & Push-Pull & \\
\hline
$L_1$:      \!\!\!\!&\!\!\!\! $\langle\psi\rangle_1$      & $\pi/4$        & $3.01/\sqrt{N}$ 
	& $0.98/\sqrt{N}$ & $1.38/\sqrt{N}$ & $2.18$ \\
$L_2$:      \!\!\!\!&\!\!\!\! $\langle\psi\rangle_2$      & $\pi/\sqrt{6}$ & $\sqrt{4\log(N/1.2)/N}$ 
	& $1.28/\sqrt{N}$ & $1.82/\sqrt{N}$ & $1.10\sqrt{\log(N/1.2)}$ \\
$L_\infty$: \!\!\!\!&\!\!\!\! $\langle\psi\rangle_\infty$ & $\pi$          & $\pi$ 
	& $2.65/\sqrt{N}$ & $3.75/\sqrt{N}$ & $O(\sqrt{N})$ \\
IQR:        \!\!\!\!&\!\!\!\! $\Delta\psi_{\rm IQR}$      & $\pi/2$        & $1.9/\sqrt{N}$ 
	& -- & -- & -- \\
\hline\hline
\end{tabular}
\caption{Phase-shift moments for the MZI (Eqs.~(\ref{eq:psi1mzi}-\ref{eq:iqrmzi})) and 3-MZI mesh (Eq.~(\ref{eq:psi3mzi})) architectures as compared to the general information-theoretic bound (Eq.~(\ref{eq:psibnd})) and the push-pull variant applicable to meshes.  We also compare the 3-MZI's moments to the (push-pull) bound to show the near-optimality of the design.}
\label{tab:t2}
\end{center}
\end{table*}

As an alternative, we consider the 3-MZI proposed in Ref.~\cite{hamerly2022asymptotically}, which consists of a standard MZI with a third 50:50 splitter that encloses the external phase shifter $\phi$ (Fig.~\ref{fig:f4}(b)).  The coupling matrix takes the form
\begin{align}
	& T_{\text{3-MZI}} = \frac{1}{2^{3/2}} \begin{bmatrix} 1 & i \\ i & 1 \end{bmatrix}
		\begin{bmatrix} e^{i\theta} & 0 \\ 0 & 1 \end{bmatrix} 
		\begin{bmatrix} 1 & i \\ i & 1 \end{bmatrix}
		\begin{bmatrix} e^{i\phi} & 0 \\ 0 & 1 \end{bmatrix}
		\begin{bmatrix} 1 & i \\ i & 1 \end{bmatrix}  \nonumber \\
	& \propto
	  	\begin{bmatrix} -\cos\bigl(\frac{\theta-\phi}{2}\bigr) + i \sin\bigl(\frac{\theta+\phi}{2}\bigr) & 
			-\sin\bigl(\frac{\theta-\phi}{2}\bigr) + i \cos\bigl(\frac{\theta+\phi}{2}\bigr) \\
			\sin\bigl(\frac{\theta-\phi}{2}\bigr) + i \cos\bigl(\frac{\theta+\phi}{2}\bigr) & 
			-\cos\bigl(\frac{\theta-\phi}{2}\bigr) - i \sin\bigl(\frac{\theta+\phi}{2}\bigr) \end{bmatrix}
\end{align}
Here, the splitting ratio is
\beq
	s_{\text{3-MZI}} = \frac{\cos\bigl(\frac{\theta-\phi}{2}\bigr) - i \sin\bigl(\frac{\theta+\phi}{2}\bigr)}{\sin\bigl(\frac{\theta-\phi}{2}\bigr) - i \cos\bigl(\frac{\theta+\phi}{2}\bigr)}
	= \frac{s_{\text{MZI}} + i}{1 + i s_{\text{MZI}}} \label{eq:s3}
\eeq
which is related to the MZI's splitting ratio by the M\"obius transformation $s \rightarrow (s+i \tan\eta)/(1+is\tan\eta)$, a consequence of the third beamsplitter in the 3-MZI (where $\eta = \pi/4$ because the splitter is 50:50).  This transformation changes the coordinate mapping $(\theta, \phi) \rightarrow s$ as illustrated in Fig.~\ref{fig:f4}(c).  The MZI's mapping resembles polar coordinates, which means that (as mentioned earlier) full programmability of the phase $\phi \in [-\pi, +\pi]$ is needed to realize all small values of $s$, the near-cross-state configurations that dominate in large meshes.  For the 3-MZI, however, the coordinates are locally Cartesian near $s = 0$.  Expanding Eq.~(\ref{eq:s3}) around the point $(\theta, \phi) = (+\pi/2, -\pi/2)$ (which maps to $s = 0$), one finds:
\beq
	s_{\text{3-MZI}} = -\frac{i}{2} (\Delta\theta + i \Delta\phi) + O(\Delta\theta^3, \Delta\phi^3) \label{eq:s3-2}
\eeq
where $\Delta\theta = \theta - \pi/2$, $\Delta\phi = \phi + \pi/2$.  
Thanks to the locally Cartesian mapping at small $s$, both $\theta$ and $\phi$ cluster about their cross-state value $\pm \pi/2$, as seen in Fig.~\ref{fig:f4}(c).  Thus, we expect to see a tightly concentrated distribution of phase shifts in the limit of large meshes, unlike in the standard MZI.

As before, we quantify this statement by calculating the moments of the phase-shift distribution following the procedure of Sec.~\ref{sec:genmzi}.  To get the moments for $\theta$ and $\phi$, we combine Eqs.~(\ref{eq:ps}) and (\ref{eq:s3-2}) to obtain the probability density for a rank-$k$ 3-MZI's phase shifters
\begin{align}
	P_k(\theta, \phi) & = \Bigl|\frac{\partial(s_{\rm re}, s_{\rm im})}{\partial(\theta, \phi)} \Bigr| P(s)
	= \frac{k}{4\pi(1+|s|^2)^{k+1}} \nonumber \\
	& \stackrel{k \gg 1}{\longrightarrow} \frac{k}{4\pi} e^{-k(\Delta\theta^2+\Delta\phi^2)/4} \label{eq:pk3}
\end{align}
This distribution is tightly centered around a mean $\theta = \pi/2$, $\phi = -\pi/2$, with a standard deviation that scales as $1/\sqrt{k}$, which is $O(1/\sqrt{N})$ for most crossings in a large mesh.  Since these Gaussians are centered about nonzero means, the moments are well approximated by their mean values
\beq
	\langle \theta \rangle_1 \approx \langle \phi \rangle_1
	\approx \langle \theta \rangle_2 \approx \langle \phi \rangle_2 \approx \pi/2
\eeq
Since the distributions are centered around nonzero values of $\theta, \phi$, the average phase shift will be $O(1)$ in this case, no improvement over the standard MZI.  However, since the variances are small, we can apply physical phase offsets at fabrication time (e.g.\ by changing the lengths or thicknesses of waveguide segments), so that the tunable phase shifters only have to realize the residual terms $\Delta\theta, \Delta\phi$ (Fig.~\ref{fig:f4}(d)), whose distributions center around zero under Eq.~(\ref{eq:pk3}).

Given the Gaussian distribution Eq.~(\ref{eq:pk3}), we have $\langle|\Delta\theta|\rangle_k = \langle|\Delta\phi|\rangle_k = 2/\sqrt{\pi k}$ and $\langle \Delta\theta^2 \rangle_k = \langle \Delta\phi^2 \rangle_k = 2/k$ for a rank-$k$ crossing (here the moments for $\Delta\theta$ and $\Delta\phi$ are the same because the Gaussian is symmetric).

Following Eq.~(\ref{eq:fsumk}), we find the average moments for $\Delta\theta$ and $\Delta\phi$ by summing over all MZI ranks.  For the $L_1$ norm, we replace the discrete sum with an integral $\sum_k \rightarrow \int{\d k}$, while the $L_2$ norm uses the standard formula for harmonic series:
\begin{align}
	\begin{array}{c} \langle \Delta\theta \rangle_{1} \\ \langle \Delta\phi \rangle_{1} \end{array} \!\!\biggr\}\,  
	& \approx \frac{2}{N^2} \int_0^N{(N-k) \frac{2}{\sqrt{\pi k}} \d k} = \frac{16}{3\sqrt{\pi N}} \label{eq:p1-3} \\
	\begin{array}{c} \langle \Delta\theta \rangle_{2} \\ \langle \Delta\phi \rangle_{2} \end{array} \!\!\biggr\}\,  
	& = \sqrt{\frac{2}{N^2}\!\sum_k {\!\frac{2(N\!-\!k)}{k}}} \approx \sqrt{\frac{4 \log(N/1.5)}{N}} \label{eq:p2-3} \\
	\begin{array}{c} \Delta\theta_{\rm IQR} \\ \Delta\phi_{\rm IQR} \end{array} \!\!\biggr\}\,  
	& \approx 1.9/\sqrt{N}
\end{align}
From this and Eqs.~(\ref{eq:psi1}-\ref{eq:psi2}), we obtain the moments of $\psi$:
\beq
	\langle \psi \rangle_1 = \frac{16}{3\sqrt{\pi N}},\ \ \ 
	\langle \psi \rangle_2 = \sqrt{\frac{4 \log(N/1.2)}{N}},\ \ \ 
	\Delta\psi_{\rm IQR} = \frac{1.9}{\sqrt{N}} \label{eq:psi3mzi}
\eeq
The terms in the logarithms above $e^{1-\gamma} \approx 1.5$ and $e^{1-\gamma-\pi/12} \approx 1.2$ come from the harmonic series Eq.~(\ref{eq:p2-3}), while the constant 1.9 is numerically evaluated.

Table~\ref{tab:t2} compares the the moments for the MZI, 3-MZI, and theoretical bound.

\subsection{Agreement with Numerics}
\label{sec:num}

\begin{figure*}[t!]
\begin{center}
\includegraphics[width=1.00\textwidth]{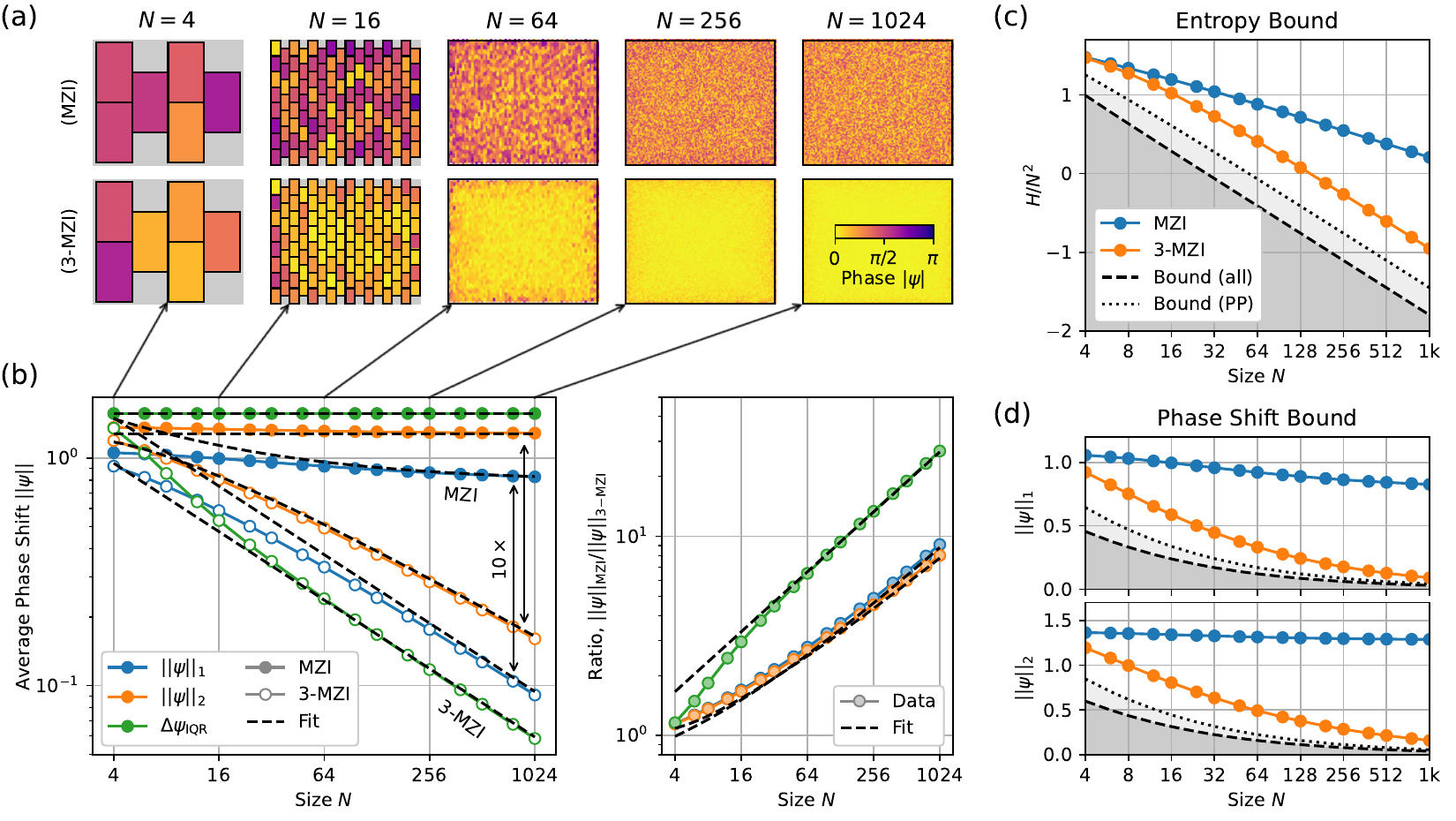}
\caption{(a) Visualization of meshes of size $N = 4$--1024, programmed to realize specific Haar-uniform sampled matrices, where MZIs are colored according to their average phase shift $|\psi| \equiv \tfrac12(|\theta| + |\phi|)$.  (b) Plot of the phase-shift moments as a function of mesh size (left) and the MZI/3-MZI ratio for each moment type, illustrating the advantage of the latter mesh type (right).  (c) Entropy per degree of freedom of the MZI and 3-MZI meshes, as compared to the information-theoretic bound.  Both generic and push-pull (PP) bounds are plotted.  (d) Phase-shift moments as compared to the bound.}
\label{fig:f5}
\end{center}
\end{figure*}

We verify these results by numerically computing the phase-shift moments of meshes programmed to an ensemble of Haar-random target matrices.  All simulations were performed using the \texttt{Meshes} package \cite{Meshes} and the code is provided in the Supplementary Material \cite{Supp}.  We program MZI and 3-MZI meshes of size up to $N = 1024$ to realize a set of target matrices (at least 10 matrices per $N$ value) uniformly sampled over the Haar measure (Haar-uniform sampling can be done by QR orthogonalization of a Gaussian random matrix).  In Fig.~\ref{fig:f5}(a), we program meshes of sizes $N = 4$, 16, 64, 256, and 1024, and plot the average phase shift per crossing, ($\tfrac12(|\theta| + |\phi|)$ for MZI, $\tfrac12(|\Delta\theta| + |\Delta\phi|)$ for 3-MZI) as a function of the position in the mesh.  For thermo-optic meshes, this plot reflects the distribution of heating power on the chip.  The average phase is highest near the edges, and lowest at the center where $\theta \approx 0$.  The 3-MZI has a lower average phase shift, particularly near the center of the mesh for large $N$.

Fig.~\ref{fig:f5}(b) plots the phase-shift moments as a function of size $N$.  Consistent with our expectations, the data show different scaling behavior, $O(1)$ and $O(1/\sqrt{N})$ for the MZI and 3-MZI, respectively.  This scaling difference leads to a 10--20$\times$ reduction in the average phase shift, depending on the figure of merit one chooses to use, for meshes of size $N \gtrsim 256$.

To compare to the information-theoretic bound, we first plot the information entropy per mode $H/N^2$ in Fig.~\ref{fig:f5}(c).  The entropy is the sum over all $\theta$, $\phi$, and phase-screen shifters
\beq
	H = \sum_k (N-k) (H_{k,\theta} + H_{k,\phi}) + N H_{\rm screen} \label{eq:htot}
\eeq
with the per-MZI entropies $H_{k,\theta} = -\int{P_k(\theta)\log P_k(\theta)\d\theta}$ and $H_{k,\phi} = -\int{P_k(\phi)\log P_k(\phi)\d\phi}$, and $H_{\rm screen} = \log(2\pi)$ due to the uniform distribution.  To derive $H$ in the large-$N$ limit, we use the $k \gg 1$ approximations to $P(\theta, \phi)$ in Eqs.~(\ref{eq:ps}, \ref{eq:pk3}) to derive the per-MZI entropies, then combine them using Eq.~(\ref{eq:htot}) to get the total entropy per mode.  The exact entropy is plotted in Fig.~\ref{fig:f5}(c), while the large-$N$ asymptotic limit is given in Table~\ref{tab:t3}.

Since the phase distribution becomes more clustered for large meshes, it is not surprising that $H/N^2$ decreases with $N$; this qualitative behavior is observed both for the observed entropy and the information-theoretic bound, which takes the form $H/N^2 \geq \tfrac12\log(2\pi e^{3/2}/N)$ for all circuits, with a stricter $H/N^2 \geq \tfrac12\log(4\pi e^{3/2}/N)$ applying to mesh-like circuits (the push-pull bound, see Sec.~\ref{sec:derbnd}).  As Fig.~\ref{fig:f5}(c) shows, the regular MZI deviates significantly from the bound and and does not even show the same scaling; on the other hand, the 3-MZI hugs the bound, with a constant difference of 0.5~nats (0.72~bits) per phase shifter.

\begin{table}[t]
\begin{center}
\begin{tabular}{r|cc}
\hline\hline
& (MZI) & (3-MZI) \\ \hline
$H_{k,\theta} = $ & $\tfrac12 \log(e^{\gamma_e+2}/k)$ & $\tfrac12\log(4\pi e/k)$ \\
$H_{k,\phi} = $   & $\log(2\pi)$                      & $\tfrac12\log(4\pi e/k)$ \\ \hline
$H/N^2 = $    & $\tfrac14 \log(4\pi^2 e^{7/2 + \gamma_e} / N)$ & $\tfrac12\log(4\pi e^{5/2}/N) \vphantom{\Bigr|}$ \\
\hline\hline
\end{tabular}
\caption{Information entropy of the phase distributions for Haar-random unitaries realized by MZI and 3-MZI mesh.  Asymptotic limit for large $N \gg 1$.}
\label{tab:t3}
\end{center}
\end{table}

Finally, Fig.~\ref{fig:f5}(d) compares the moments for the MZI (Sec.~\ref{sec:mzi}), 3-MZI (Sec.~\ref{sec:3mzi}), and lower bound (Sec.~\ref{sec:bound}).  As we already knew, the MZI and 3-MZI have different scalings for the moments with respect to $N$, with the 3-MZI's norms decreasing as $O(1/\sqrt{N})$ while the MZI's norms asymptote to a constant.  What is remarkable here is that the 3-MZI does not only enjoy the same scaling as the bound (up to a logarithmic term), but it reaches the bound to just over a factor of $2\times$ for realistic mesh sizes: for $N = 256$, the $L_1$ [resp.\ $L_2$] norm is only about $2.2\times$ [resp.\ $2.6\times$] the theoretical minimum, so the 3-MZI mapping is very close to optimum.  In the $N \rightarrow \infty$ limit, the deviation from optimality is obtained by dividing Eq.~(\ref{eq:psi3mzi}) by Eq.~(\ref{eq:psibnd}):
\beq
	\frac{\lVert\psi\rVert_1}{\lVert\psi\rVert_{1,\rm min}} =
		\frac{16 e^{1/4}}{3\pi} \approx 2.20,\ \ \ \ \ 
	\frac{\lVert\psi\rVert_2}{\lVert\psi\rVert_{2,\rm min}} \rightarrow
		\sqrt{\frac{2 \log(N/1.5)}{e^{1/2}}}
\eeq
Technically, the gap for the $L_2$ norm diverges, but it does so only logarithmically, and for reasonable mesh sizes $N \lesssim 10^3$, it is in the range of 2--3$\times$.

\section{Saturating the Bound: Gaussian Matrices with Crossbar Meshes}

\begin{figure}[b!]
\begin{center}
\includegraphics[width=1.00\columnwidth]{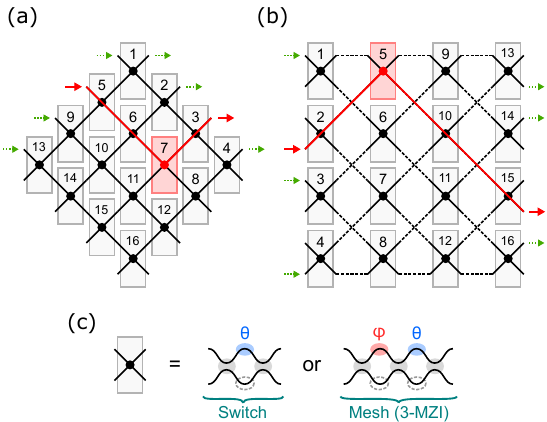}
\caption{Crossbar-based non-unitary mesh architectures: (a) Diamond and (b) PILOSS.  I/O ports are labeled with arrows.  (c) Crossing type used for switch (left) and 3-MZI mesh (right)}
\label{fig:f6}
\end{center}
\end{figure}

The analysis above was restricted to Haar-uniform unitary matrices, where the theory of meshes is most developed.  However, most matrices used in information processing are non-unitary, and cannot be represented by a Reck or Clements mesh.  As this section shows, we can also construct phase-efficient meshes for non-unitary matrices.  In fact, for the right target distribution (Gaussian random matrices with small entries), the 3-MZI mesh {\it exactly reaches} the information-theoretic bound, without any factor of $2\times$, making the design provably optimal.

\subsection{Diamond and PILOSS Crossbar Meshes}

A fully general (i.e.\ non-unitary) $N\times N$ beamsplitter mesh can be realized as an analog generalization of a crossbar switch.  Two common switch architectures, the diamond \cite{Blass1960, Mosca2002, Taballione2019} and path-independent loss (PILOSS) \cite{suzuki2014ultra, Suzuki2018}, are shown in Fig.~\ref{fig:f6}(a-b).  Depending on the crossing employed, Fig.~\ref{fig:f6}(c), the crossbar will function as either a digital switch or an analog matrix multiplier.  Of the $2N$ input/output (I/O) ports of these crossbars, only $N$ are used in the circuit (the remainder serving as auxiliary ``dark ports'' for unwanted power) and every pair of connected I/O ports maps to a unique crossing in the mesh (red in Fig.~\ref{fig:f6}).

To program the crossbar as a {\it switch}, first we initialize all crossings (MZIs) to the CROSS state $s = 0$, so that power passes straight along the diagonals.  Next, for each input/output pair that we desire to connect, we trace the light paths and set the corresponding crossing to the BAR state $s = \infty$.

To program the crossbar to realize a general, nonunitary {\it matrix}, employ a similar procedure, equivalent to the ``direct'' method of Ref.~\cite{hamerly2022stability}:
\begin{enumerate}
	\item Initialize all 3-MZI crossings to the CROSS state.
	\item Following the order shown in Fig.~\ref{fig:f6}(a-b):
	\begin{enumerate}
		\item Trace the diagonals to find the inputs and outputs $\text{in}_j$, $\text{out}_i$ corresponding to this crossing (red in figure).
		\item Adjust the phase shifts of each 3-MZI to set the measure matrix element $M_{ij}$ to its target value.
	\end{enumerate}
	\item (PILOSS mesh only) Repeat step 2 until convergence.
\end{enumerate}

While other non-unitary mesh structures exist, e.g.\ based on the singular value decomposition \cite{miller2013self}, the crossbar-based architectures studied here have the advantage of a uniform layout and straightforward programming procedure, in addition to saturating the information-theoretic bound as we show later in this section.

\subsection{Entropy Bound for Random Matrices}

Here, we consider a distribution of $N\times N$ target matrices $M$, where each entry $M_{ij}$ is i.i.d.\ with zero mean and a standard deviation $\sigma$ in both real and imaginary part:
\beq
	\langle M_{ij} \rangle = 0,\ \ \ 
	\langle \text{Re}[M_{ij}]^2 \rangle = \langle \text{Im}[M_{ij}]^2 \rangle = \sigma^2,\ \ \ 
	\langle |M_{ij}|^2 \rangle = 2\sigma^2
\eeq
The entropy is minimized with a Gaussian distribution with $\tfrac12\log(2\pi e \sigma^2)$ per mode, and with $2N^2$ modes (counting both real and imaginary parts of each $M_{ij}$), we get $H_M \geq N^2 \log(2\pi e\sigma^2)$.  However, $M$ is embedded in a larger matrix $U$, which realizes the full mesh with all dark ports included, and we need to find $H_U$ in order to get the entropy bound for $\langle \psi \rangle$.  Since $U$ contains $M$, clearly $\lVert \Delta U \rVert \geq \lVert \Delta M \rVert$ for any perturbation $\Delta M$, which implies that $|\partial(U)/\partial(M)| \geq 1$ and at a minimum $H_U \geq H_M$.

\begin{figure*}[htbp]
\begin{center}
\includegraphics[width=1.00\textwidth]{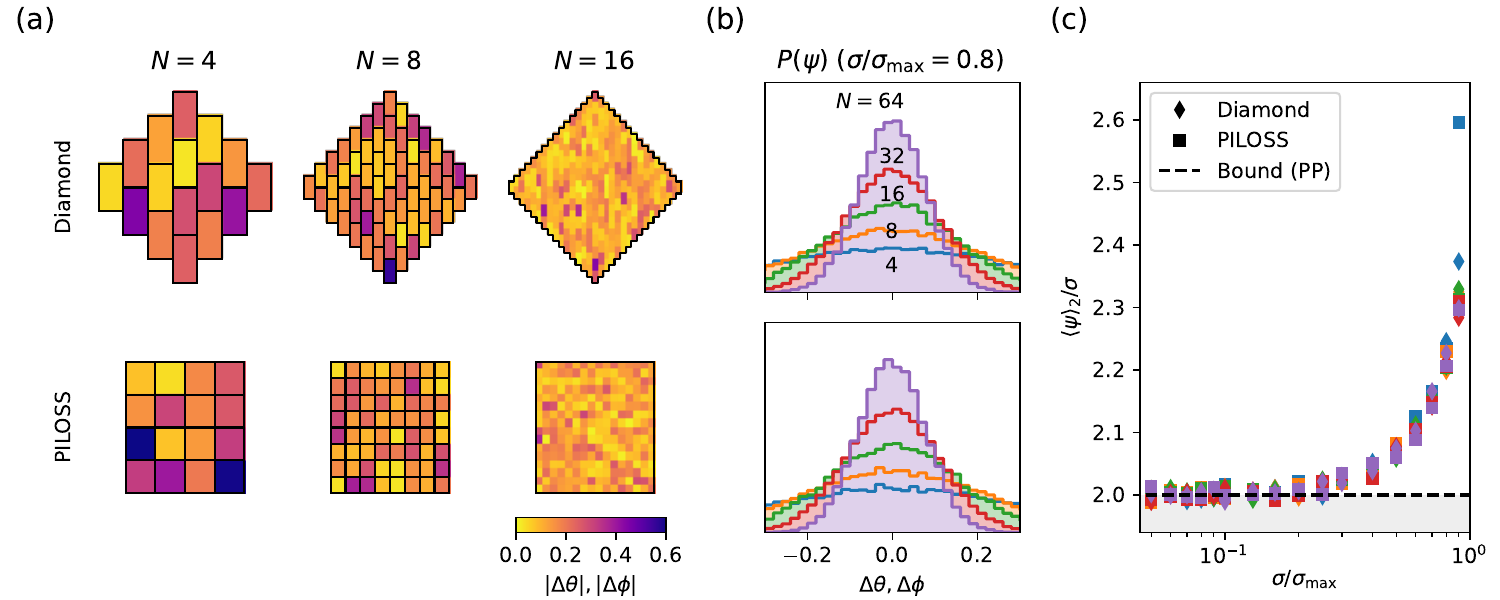}
\caption{Non-unitary meshes with Gaussian random target matrices.  (a) Phase shift $\Delta\theta, \Delta\phi$ for diamond and PILOSS meshes, with target matrices randomly sampled with $\sigma/\sigma_{\rm max} = 0.8$.  (b) Phase shift distribution as a function of mesh size, showing a variance reduction consistent with $\langle\psi\rangle \propto 1/\sqrt{N}$.  (c) $L_2$-average phase shift as a function of $\sigma/\sigma_{\rm max}$, compared to the (push-pull) information-theoretic bound $\langle \psi\rangle_2 \geq 2\sigma$.}
\label{fig:f7}
\end{center}
\end{figure*}

However, we can obtain a stricter bound that holds in the limit of small $\sigma$.  Consider an embedding $g: \mathbb{C}^{N^2} \rightarrow U(2N)$ into a $(2N)\times (2N)$ space as in Fig.~\ref{fig:f6}(a), where the first $N$ ports are for I/O and the rest are auxiliary.  Up to a transform that does not affect the entropy,
the unitary is $g(0) = [[0, 1_{N\times N}], [1_{N\times N}, 0]]$.  %
Taking $\epsilon$ to be a small parameter and linearizing, the map $g$ takes the following form:
\beq
	g(\epsilon M) = \begin{bmatrix} \epsilon M & 1 + \epsilon A \\ 1 + \epsilon B & \epsilon C \end{bmatrix} + O(\epsilon^2)
\eeq
where $A, B, C$ matrices that depend on $M$.  Unitarity $U^\dagger U = 1$ sets the constraints $A = B = 0$, $C = M^\dagger$, so the embedded $U$ takes the form:
\beq
	U \approx \begin{bmatrix} M & 1 \\ 1 & M^\dagger \end{bmatrix}
\eeq
This embedding has two $M$ blocks, so $\lVert \Delta U \rVert = \sqrt{2} \lVert \Delta M \rVert$, implying $|\partial(U)/\partial(M)| \geq 2^{N^2}$.  Therefore, the entropy $H_U$ is larger by $\tfrac12\log(2)$ per mode.  This, combined with the extra factor of $\sqrt{2}$ for push-pull MZI circuits (discussion at end of Sec.~\ref{sec:derbnd}), gives us the following lower bounds:
\begin{align}
	H_U & \geq H_M + N^2 \log(2) = N^2 \log(4\pi e \sigma^2) \\
	H_\psi & \geq H_U + N^2 \log(2) = N^2 \log(8\pi e \sigma^2)
\end{align}
Using the Lagrange multiplier technique following Eq.~(\ref{eq:psibnd0}) (but substituting $e^{H/N} \rightarrow e^{H/2N}$ because of the larger number of phase shifters), we obtain the following lower bound for the $L_2$ moment:
\beq
	\langle\psi\rangle_2 \geq 2\sigma \label{eq:psibnd_nu}
\eeq
How well does a 3-MZI based diamond or PILOSS mesh compare?  In the weak-scattering limit $|M| \ll 1$, the crossings are all very close to the CROSS state, and each matrix element is (up to a constant phase) approximately set by the $T_{11}$ element of its corresponding crossing:
\beq
	M_{ij} \approx (T_{11})_{ij} = \frac{e^{i\pi/4} (\Delta\theta+i\Delta\phi)}{2}
\eeq
which gives $\langle\theta\rangle_2 = \langle\phi\rangle_2 = 2\sigma$ for the normally-distributed $M_{ij}$ considered here.  Therefore, we expect the 3-MZI mesh to asymptotically approach the information-theoretic bound for Gaussian distributions in the weak-scattering limit.

To show this numerically, Fig.~\ref{fig:f7} shows the phase-shift statistics for the two non-unitary meshes.  The result depends on $\sigma$, which must be properly normalized to ensure energy conservation.  In a passive optical circuit, this requires that $N$ be contractive $|M| < 1$, i.e.\ all singular values $s_i$ should be less than unity.  In the large-$N$ limit, Marchenko-Pastur theory applies \cite{marchenko1967distribution}, and the distribution of singular values is given by $P(s) = \sqrt{8\sigma^2 N - s^2}/2\pi\sigma^2 N$, which is upper-bounded by $s_{\rm max} = 2\sqrt{2N}\sigma$.  Defining $\sigma_{\rm max} = 1/(2\sqrt{2N})$, $M$ is contractive in the large-$N$ limit whenever $\sigma/\sigma_{\rm max} < 1$.  

In Fig.~\ref{fig:f7}(a), we plot the phase shifts ($\Delta\theta$, $\Delta\phi$) for randomly chosen Gaussian matrices at $\sigma/\sigma_{\rm max} = 0.8$.  As the mesh size $N$ becomes larger, the individual phase shifts become smaller, a trend that is confirmed by plotting the distributions in Fig.~\ref{fig:f7}(b).  Since $\psi \propto \sigma_{\rm max} \propto 1/\sqrt{N}$, we find the same scaling law as for the unitary mesh.

Finally, Fig.~\ref{fig:f7}(c) plots the numerically computed ratio $\langle\psi\rangle_2/\sigma$, for both diamond and PILOSS architectures of a range of mesh sizes, as a function of $\sigma/\sigma_{\rm max}$.  For almost all cases, the 3-MZI mesh is within 10\% of the bound $\langle\psi_2\rangle/\sigma = 2$ (Eq.~(\ref{eq:psibnd_nu})), while the bound is almost exactly satisfied for $\sigma/\sigma_{\rm max} \lesssim 0.3$.

\section{$L_{\infty}$-constrained DNN Training}

\begin{figure}[b!]
\begin{center}
\includegraphics[width=1.00\columnwidth]{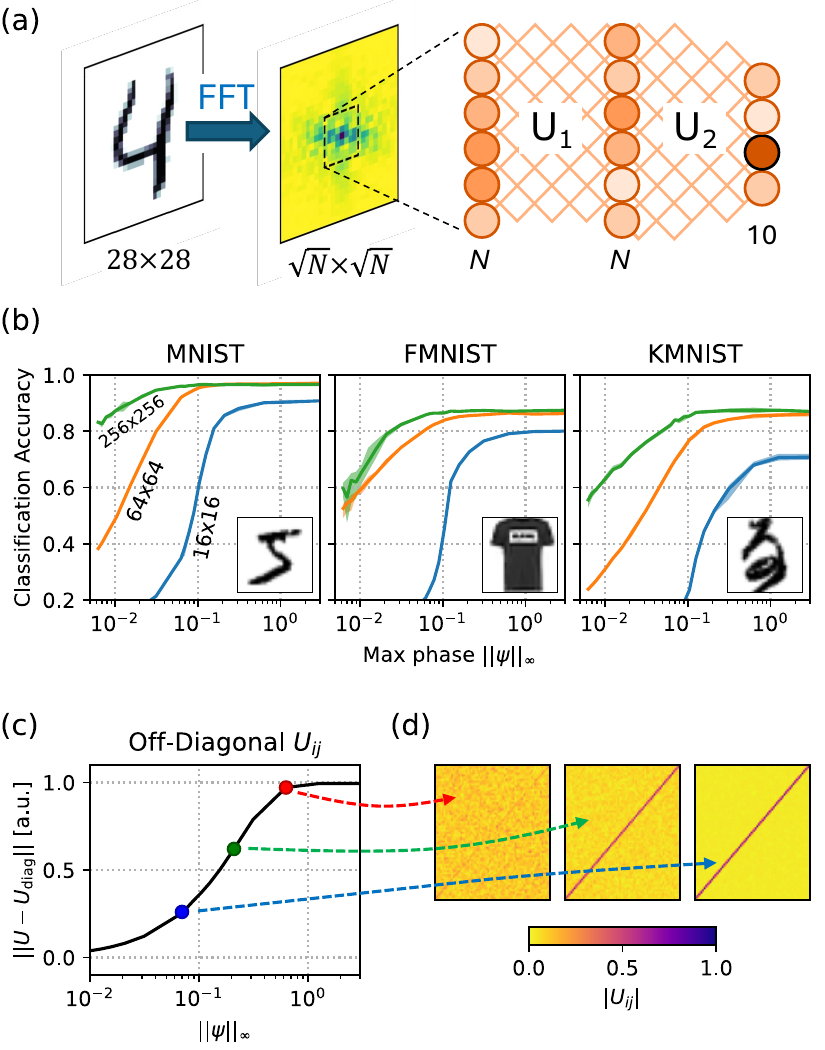}
\caption{Phase-constrained ONN training.  (a) Architecture of ONN.  The FFT and cropping are preprocessing steps, while the $N\times N$ trainable layers $U_1$ and $U_2$ are implemented with Clements meshes.  (b) MNIST, FMNIST, and KMNIST classification accuracy as a function of layer size $N$ and maximum phase $\lVert\psi\rVert_\infty$.  (c) Fraction of matrix norm of $U_1$ contributed by off-diagonal elements, as a function of $\lVert\psi\rVert_\infty$.  MNIST, $N = 64$; the curve for $U_2$ is similar.  (d) Representative matrices from (c), showing the increased clustering about the anti-diagonal when $\lVert\psi\rVert_\infty$ is small.}
\label{fig:f8}
\end{center}
\end{figure}

For many applications, the each phase shifter has a ``hard'' upper limit on the phase shift achievable, i.e.\ an $L_{\infty}$ constraint.  As we discussed previously in the context of {\it universal} meshes, the $L_{\infty}$ norm is sensitive to {\it worst-case} statistics and does not enjoy a $1/\sqrt{N}$ scaling like the other norms.  However, applications in optical deep learning do not require full universality but only a high degree of connectivity.  Therefore, it is worth studying empirically whether optical neural networks can be trained under an $L_{\infty}$ constraint, to achieve accuracy comparable to unconstrained models.

Here, we consider the optical neural network in Fig.~\ref{fig:f8}(a): an image is Fourier transformed, clipped to size $\sqrt{N}\times \sqrt{N}$ ($N$ neurons), and sent into a sequence of two Clements meshes of size $N\times N$, with an electro-optic nonlinear layer between the meshes.  Details can be found in Refs.~\cite{hamerly2022asymptotically, williamson2019reprogrammable}.  To implement the $L_{\infty}$ constraint in training, we must impose the conditions $|\Delta\theta| \leq \lVert\psi\rVert_{\infty}$, $|\Delta\phi| \leq \lVert\psi\rVert_{\infty}$ on all phase shifters.  However, for added robustness to fabrication phase error, we will instead train subject to the stricter condition $\Delta\theta^2 + \Delta\phi^2 \leq \lVert\psi\rVert_{\infty}^2$.  For small $\lVert\psi\rVert_{\infty}$, this is equivalent to upper-bounding the splitting ratio $|s| \leq \lVert\psi\rVert_{\infty}/2$.  Note the similarity to the procedure taken in Ref.~\cite{vadlamani2023transferable}, where one imposes a {\it lower-bound} on $|s|$ in training, in order to make the trained models more resilient to hardware errors.  The model is trained on MNIST \cite{lecun1998gradient}, Fashion MNIST \cite{xiao2017fashion}, and KMNIST (Japanese Kanji) \cite{clanuwat2018deep}, using the Adam optimizer and mean-squared-error loss, for a range of of $\lVert\psi\rVert_{\infty}$ and mesh sizes $N = 16, 64$, and $256$ to determine the effect of both model size and phase constraint on the neural network accuracy.

Fig.~\ref{fig:f8}(b) plots the classification accuracy for all three problems as a function of $\lVert\psi\rVert_{\infty}$ and the network size.  In a universal mesh, $\lVert\psi\rVert_{\infty} = \pi$, and the network trains to its maximum accuracy.  Imposing the $L_{\infty}$ constraint, we see a degradation in accuracy when $\lVert\psi\rVert_\infty$ is lowered below a threshold value, roughly 0.5~rad, 0.2~rad, and 0.08~rad for $N = 16$, 64, and 256, respectively.  The dependence of this threshold on $N$ is consistent with the $O(1/\sqrt{N})$ scaling observed in the 3-MZI and the information-theoretic bound, where larger meshes require smaller average phase shift.  The required $\lVert\psi\rVert_{\infty}$ is 1-2 orders of magnitude lower than the value required for a universal mesh, and would be even lower for mesh sizes $N > 256$.

If $\lVert\psi\rVert_{\infty}$ is made too small, the crossings on the mesh scatter too little light, and the matrix begins to approximate an anti-diagonal $M_{ij} \sim \delta_{i,N+1-j}$.  This can be shown by removing this part of the matrix and calculating the norm $\lVert U - U_{\rm diag}\rVert / \lVert U \rVert$, of all elements off the (anti-)diagonal, Fig.~\ref{fig:f8}(c).  As $\lVert\psi\rVert_{\infty}$ decreases, this norm goes to zero, suggesting that the off-diagonal matrix elements become very small, and the matrix becomes close to an anti-diagonal, Fig.~\ref{fig:f8}(d).  This limits the expressivity and connectivity of the mesh, which in turn prevents the neural network achieving a high accuracy.

\section{Conclusion}

It is well understood that scaling up photonic circuits to the the very-large-scale regime will require device-level improvements, e.g.\ reducing the power and footprint of MZIs and phase shifters.  The point of this paper is that system- and architecture-level choices can have just as much impact as improved devices.  We have shown that use of the recently-proposed 3-MZI mesh (which already enjoys advantages in error robustness \cite{hamerly2022asymptotically}) can reduce the average per-MZI phase shift from $O(1)$ to $O(1/\sqrt{N})$ in universal meshes, an advantage of approximately 10--$20\times$ in near-term hardware.  But to temper this optimism, we have also derived a lower bound on the average phase shift of a mesh, rooted in information theory, that also scales as $O(1/\sqrt{N})$, so future scaling improvements will not be possible.  In fact, the 3-MZI approaches this bound to within a factor of 2--$3\times$, and is therefore near-optimal.  

Moreover, for non-unitary matrix distributions, we derive a similar bound, and show that crossbar-based meshes (specifically diamond \cite{Blass1960, Mosca2002, Taballione2019} and PILOSS \cite{suzuki2014ultra, Suzuki2018}), with a 3-MZI crossing unit, saturate the bound when the matrix elements are drawn from a Gaussian distribution.  As a side note, we believe that these crossbar meshes are relatively overlooked in a community that tends to focus on unitary matrices, and due to their straightforward programmability, they may be a better solution for most applications when compared to the more well-known SVD architecture \cite{miller2013self}.

Finally, we studied the training of optical neural networks under a ``hard'' $L_\infty$ constraint.  While this constraint breaks the mesh's universality, we are still able to train models on MNIST, FMNIST, and KMNIST to canonical accuracy even when the maximum phase shift is limited to a small fraction of a radian, a factor of 10--$20\times$ less than is required in a universal mesh.  $L_\infty$-constrained training may prove crucial in enabling mesh-based ONNs to scale up in fast platforms like LiNbO$_3$ \cite{wu2019fabrication} or piezo-optomechanics \cite{dong2022high} which suffer from severe $V_\pi L$ tradeoffs.

Several open questions merit future work.  First, the information-theoretic bound derived in this paper assumes that the matrix space ($U(N)$ for unitaries, $\mathbb{C}^{N^2}$ for non-unitaries) and phase-shift space have same dimension.  It may be nontrivial to extend this logic to overcomplete bases where there are more phase shifters than needed, as this would entail a projective mapping $f: \vec{\psi} \rightarrow U$ with a singular Jacobian, which breaks several assumptions made in our derivation.  It is also worth exploring the average phase-shift statistics of MPLC \cite{tanomura2020robust}, and FFT \cite{Pastor2021}, and other \cite{fldzhyan2020optimal} architectures, which are not MZI-based and are thus subject to the less strict general (non-push-pull) bound, allowing in principle a $\sqrt{2}\times$ reduction in phase shift.  The use of continuous systems, like multimode waveguides \cite{Larocque2021, larocque2023piezoelectric, lopez2021arbitrary, onodera2024scaling}, should also be explored, and since this breaks the lumped-element phase-shifter assumptions made here, the theoretical framework may be very different.  Finally, hardware imperfections (specifically phase-shifter errors) will play a role in any realistic mesh \cite{fang2019design}; the compatibility of hardware error correction \cite{bandyopadhyay2021hardware, hamerly2022accurate} with low average-phase-shift meshes should also be explored.

\section*{Appendix}

\setcounter{section}{0}
\renewcommand{\thesection}{\Alph{section}}

%

\section{Moments $\langle \theta\rangle_1$, $\langle \theta\rangle_2$, $\Delta\theta_{\rm IQR}$ for the MZI mesh}
\label{sec:mzith}

In Sec.~\ref{sec:mzi}, we established that the average phase shift for the MZI mesh is always $O(1)$, being dominated by the external phase shifters, for which $\phi \in [-\pi, +\pi]$ is uniformly distributed.  Since $\theta \ll 1$, the $\theta$ phase shifters will introduce only a minor $O(1/\sqrt{N})$ correction to this average, which we ignored for purposes of brevity.  For completeness, here we derive the moments for the $\theta$ phase shifts.

Exact, closed-form solutions exist for $\langle \theta \rangle_1$, $\langle \theta \rangle_2$, and $\Delta\theta_{\rm IQR}$ for each MZI rank $k$.  However, since we are interested in the asymptotic behavior of large meshes where the majority of MZIs have rank $k \gg 1$, we can use the Gaussian approximation $P_k(\theta) \approx (k\theta/2) e^{-k\theta^2/4}$ to get a good approximate formula for the moments.  Here, we have $\langle |\theta| \rangle_k = \sqrt{\pi/k}$ and $\langle \theta^2 \rangle_k = 4/k$ for a single MZI.  To obtain the average quantity over the whole mesh, we must sum over all values of $k$ and divide by $N(N-1)/2$, i.e.\ the number of MZIs:
\beq
	\langle \ldots \rangle = \frac{2}{N(N-1)} \sum_k (N-k) \langle \ldots \rangle_k
\eeq
We substitute the discrete sum over $k$ with integral $\sum_k \rightarrow \int{\d k}$, giving:
\begin{align}
	\langle \theta \rangle_1 & \approx \frac{2}{N^2} \int_0^N {(N-k) \sqrt{\pi/k}\,\d k} = \frac{8}{3}\sqrt{\pi/N} \\
	(\langle \theta \rangle_2)^2 & \approx \frac{2}{N^2} \int^N {(N-k)(4/k)\d k} \rightarrow \frac{8 \log(N/N_0)}{N} \label{eq:q2}
\end{align}
The integral in Eq.~(\ref{eq:q2}) technically diverges at $k = 0$; however, for these small values of $k$, approximating the discrete sum by an integral was not accurate to begin with.  So we truncate these divergent terms, leading to the $\log(N/N_0)/N$ form given above.  The constant $N_0 \approx 2$ is determined by matching to numerical data.

We can also obtain an analytic form for the total probability density $P(\theta) = \sum_k (N-k) P_k(\theta)$ by approximating the sum as an integral:
\begin{align}
	P(\theta) & \approx \int_0^N{(N-k)P_k(\theta)} = \frac{\d}{\d x} F(N x^2/8), \nonumber \\
	&  
	\begin{dcases}
		F(y)\, = \frac{1-2y+2y^2-e^{-2y}}{2y^2} \\
		F'(y) = \frac{2e^{-y} (y \cosh(y) - \sinh(y))}{y^3}
	\end{dcases} \label{eq:ptmzi}
\end{align}
From this, we find that the IQR of $P(\theta)$ lies between $\theta = 1.93/\sqrt{N}$ and $5.22/\sqrt{N}$, so 
\beq
	\Delta\theta_{\text{IQR}} = 3.3/\sqrt{N}.
\eeq

We can improve the $\theta$ figures slightly by adding an offset phase to each MZI $\bar{\theta} = \theta - \theta_0$, in order to center the distribution at $\bar{\theta} = 0$, as explained in the main text.  To minimize the $L_1$ norm, this offset should be the median $\theta_0 = 2\sqrt{\log(2)/k} \approx 1.67/\sqrt{k}$, while to minimize the $L_2$ norm, one picks the mean $\theta_0 = \sqrt{\pi/k} \approx 1.77/\sqrt{k}$.  The moments are reduced by constant factors:
\begin{align}
	\langle \bar{\theta} \rangle_1 / \langle\theta\rangle_1
	& = 1 - \text{erf}(\log(\sqrt{2})) + 2\sqrt{\log(2)/\pi}
		\approx 0.42 \\
	\langle \bar{\theta} \rangle_2 / \langle\theta\rangle_2
	& = \sqrt{1-\pi/4} \approx 0.46
\end{align}
Note that all of these averages $\langle \theta\rangle_1$, $\langle \theta\rangle_2$, and $\Delta\theta_{\rm IQR}$ scale as $O(1/\sqrt{N})$, while the external phase shifters will have moments that are $O(1)$.  Therefore, in the limit of large $N$, the $\theta$ phase shifters can be safely ignored when calculating the average phase shift.


\begin{table}[tbp]
\begin{center}
\begin{tabular}{c|ccc}
\hline\hline
Type 
	& $\langle\psi\rangle_1$ & $\langle\psi\rangle_2\vphantom{\Bigr|}$ & $\psi_{\text{IQR}}$ \\
\hline
MZI 
	& $\pi/2 \!+\! O(\frac{1}{\sqrt{N}})$ & $\frac{\pi}{\sqrt{6}} \!+\! O(\frac{1}{N})\vphantom{\Bigr|}$ & $\pi/2$ \\
3-MZI 
	& $\tfrac{16}{3}/\sqrt{\pi N}$ & $\sqrt{\frac{4\log(N/N''_0)}{N}}\vphantom{\Bigr|}$ & $1.9/\sqrt{N}$ \\ 
\hline
Ratio$^*$ 
	& $0.47 \!+\! 0.26 \sqrt{N}$ & $0.64\sqrt{N/\log(N)}\vphantom{\Bigr|}$ & $0.83\sqrt{N}$ \\
\hline\hline
\end{tabular}
\caption{List of phase-shifter figures of merit $\langle\psi\rangle_1$, $\langle\psi\rangle_2$, and $\psi_{\text{IQR}}$ for the MZI and 3-MZI crossing types.  $^*$Ratio refers to $(\cdot)_{\text{MZI}} / (\cdot)_{\text{3-MZI}}$.}
\label{tab:t4}
\end{center}
\end{table}


%






\bibliography{PaperRefs}{}
\bibliographystyle{IEEEtranN}

\end{document}